\documentclass[11pt,twoside]{article}

\usepackage{asp2006}
\usepackage{epsf}
\usepackage{epsfig}
\usepackage{lscape}

\newcommand{\CR}{\mathrm{CR}}
\newcommand{\therm}{\mathrm{th}}

\begin{document}
\title{Cosmic ray feedback in hydrodynamical simulations of galaxy and galaxy cluster formation}
\author{C. Pfrommer$^{1}$, V. Springel$^2$, M. Jubelgas$^2$,  T.A. En{\ss}lin$^2$}
\affil{1) Canadian Institute for Theoretical Astrophysics, Canada\\
2) Max-Planck Institut f{\"u}r Astrophysik, Germany} 

\begin{abstract}
It is well known that cosmic rays (CRs) contribute significantly to the
pressure of the interstellar medium in our own Galaxy, suggesting that they may
play an important role in regulating star formation during the formation and
evolution of galaxies. We will present a novel numerical treatment of the
physics of CRs and its implementation in the parallel smoothed particle
hydrodynamics (SPH) code GADGET-2. In our methodology, the non-thermal CR
population is treated self-consistently in order to assess its dynamical impact
on the thermal gas as well as other implications on cosmological observables.
In simulations of galaxy formation, we find that CRs can significantly reduce
the star formation efficiencies of small galaxies. This effect becomes
progressively stronger towards low mass scales.  In cosmological simulations of
the formation of dwarf galaxies at high redshift, we find that the total
mass-to-light ratio of small halos and the faint-end of the luminosity function
are affected.  In high resolution simulations of galaxy clusters, we find lower
contributions of CR pressure, due to the smaller CR injection efficiencies at
low Mach number flow shocks inside halos, and the softer adiabatic index of
CRs, which disfavours them when a composite of thermal gas and CRs is
adiabatically compressed. Within cool core regions, the CR pressure reaches
equipartition with the thermal pressure leading to an enhanced compressibility
of the central intra-cluster medium, an effect that increases the central
density and pressure of the gas.  While the X-ray luminosity in low mass cool
core clusters is boosted, the integrated Sunyaev-Zel'dovich effect is only
slightly changed. The resolved Sunyaev-Zel'dovich maps, however, show a larger
variation with an increased central flux decrement.
\end{abstract}

\section{Motivation}

The interstellar medium (ISM) of galaxies has an energy budget composed both of
thermal and non-thermal components. The non-thermal components which are
magnetic fields and cosmic rays (CRs) are known to contribute a significant
part of the energy and pressure to the ISM.  CRs behave quite differently
compared to the thermal gas. Their equation of state is softer, they are able
to propagate over macroscopic distances, and their energy loss time-scales are
typically larger than the thermal ones. Furthermore, roughly half of the
particle's energy losses are due to Coulomb and ionization interactions and
thereby heat the thermal gas. Therefore, CR populations are an important
galactic reservoir for energy from supernova explosions, and thereby help to
maintain dynamical feedback of star formation for periods longer than thermal
gas physics alone would permit. In contrast, it is unknown how much pressure
support is provided by CRs to the thermal plasma of clusters of galaxies.

\section{CR modified galaxy formation and cluster observables}

\begin{figure}
  \begin{minipage}[t]{0.32\textwidth}
    \begin{center}
    $10^{10} \, M_\odot$      
    \end{center}
    \psfig{figure=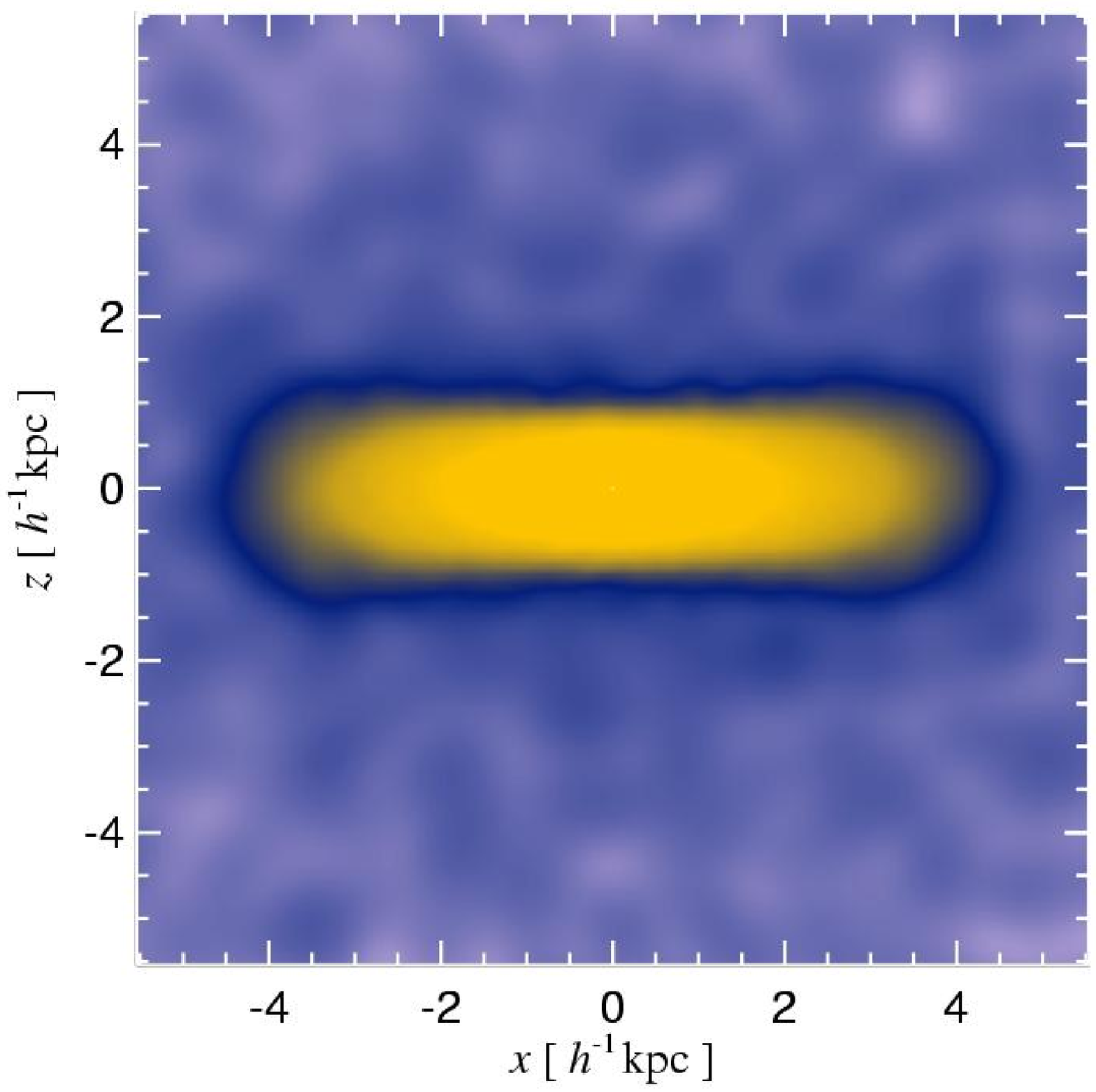,width=\textwidth}    
    \psfig{figure=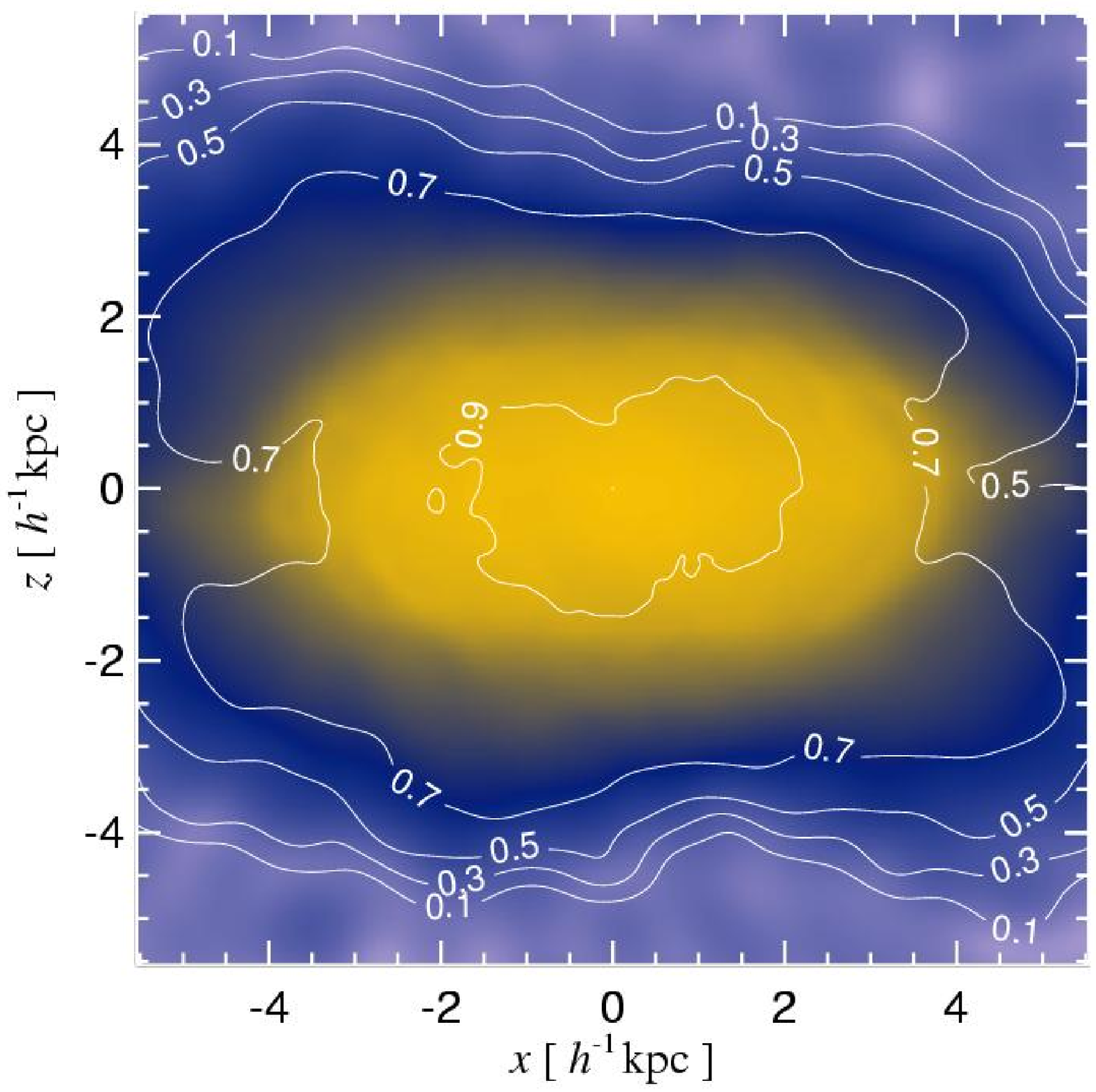,width=\textwidth}    
    \psfig{figure=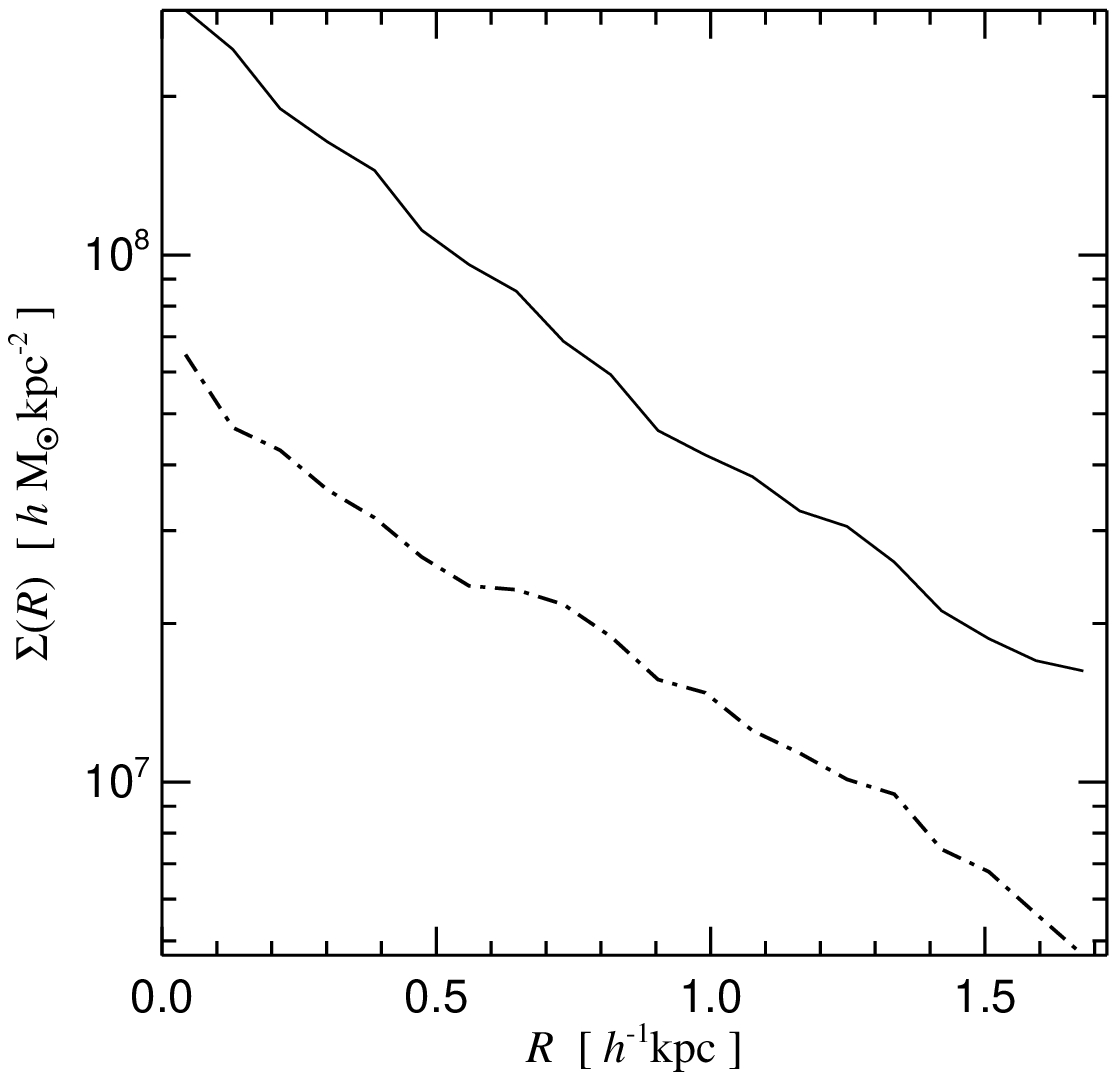,width=\textwidth}    
    \psfig{figure=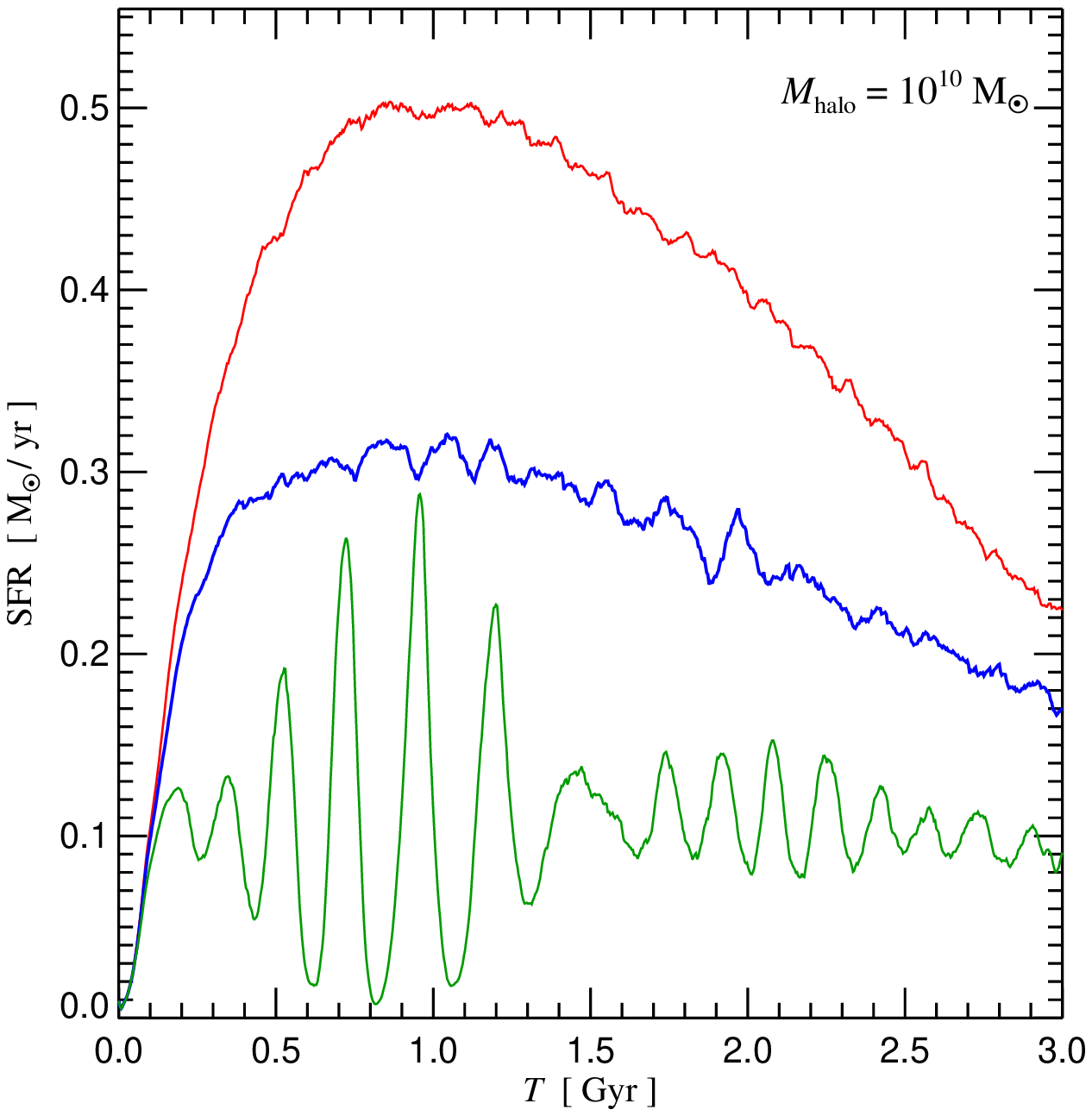,width=\textwidth}    
  \end{minipage}
  \begin{minipage}[t]{0.32\textwidth}
    \begin{center}
    $10^{11} \, M_\odot$      
    \end{center}
    \psfig{figure=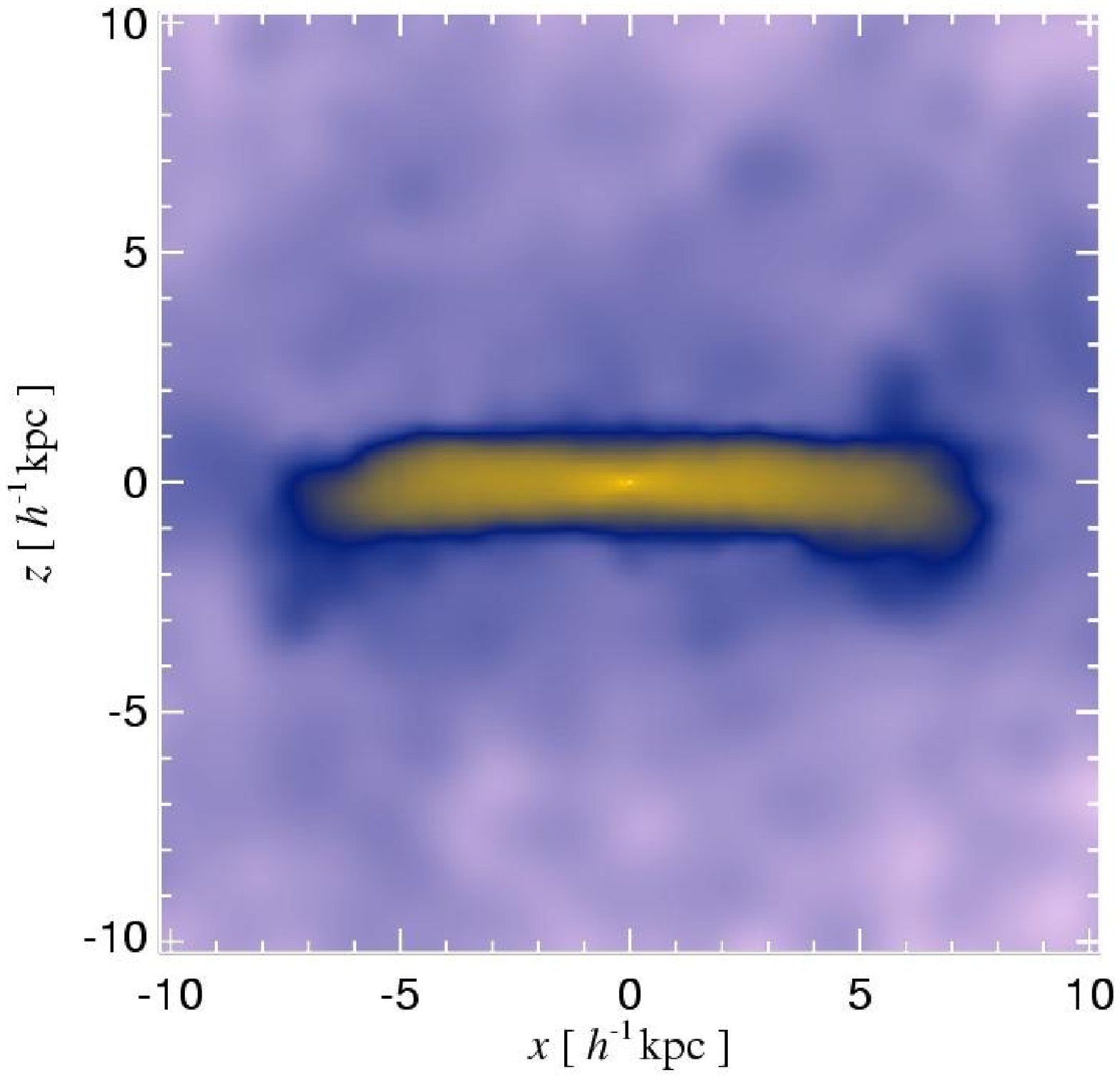,width=\textwidth}    
    \psfig{figure=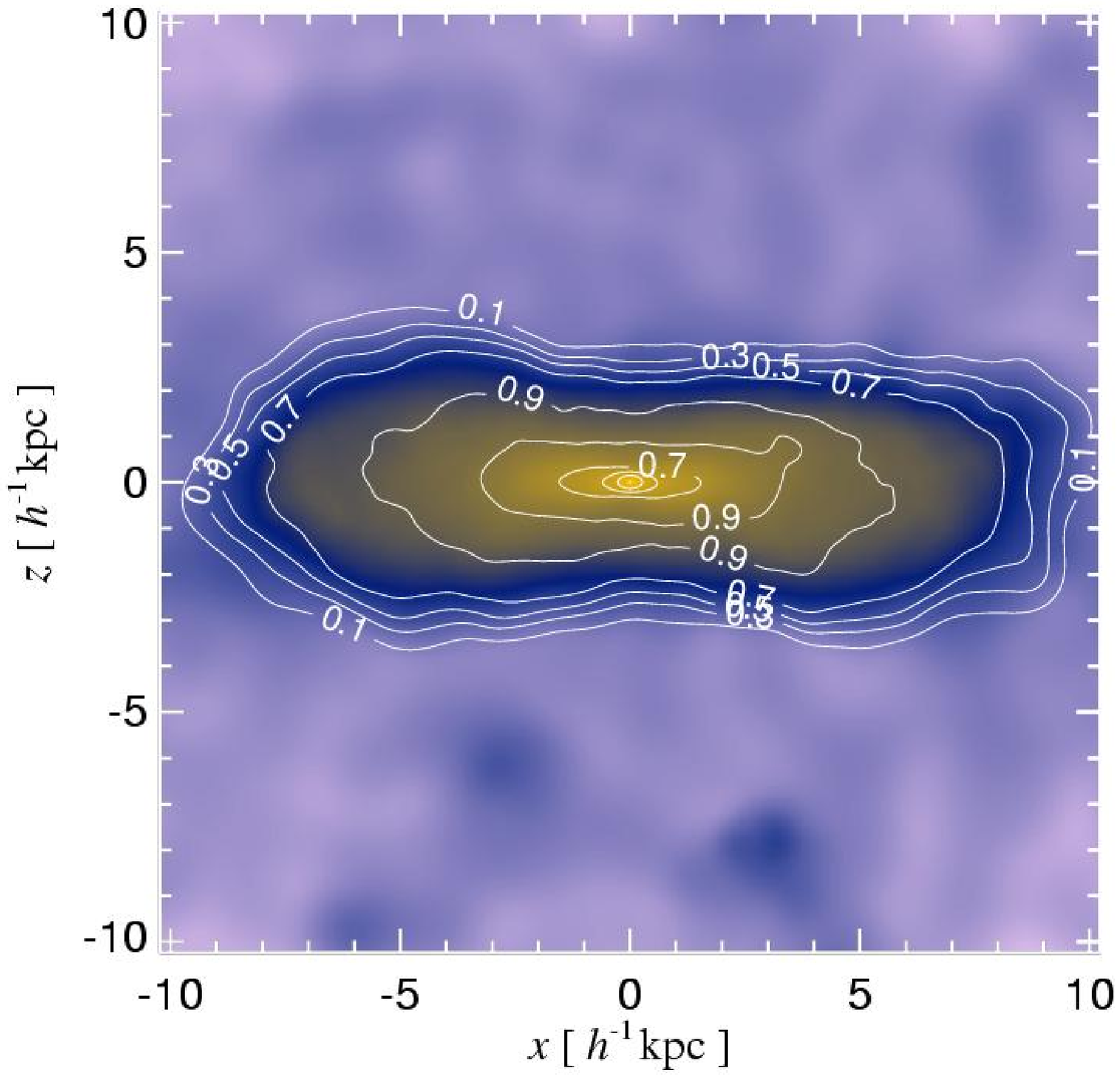,width=\textwidth}    
    \psfig{figure=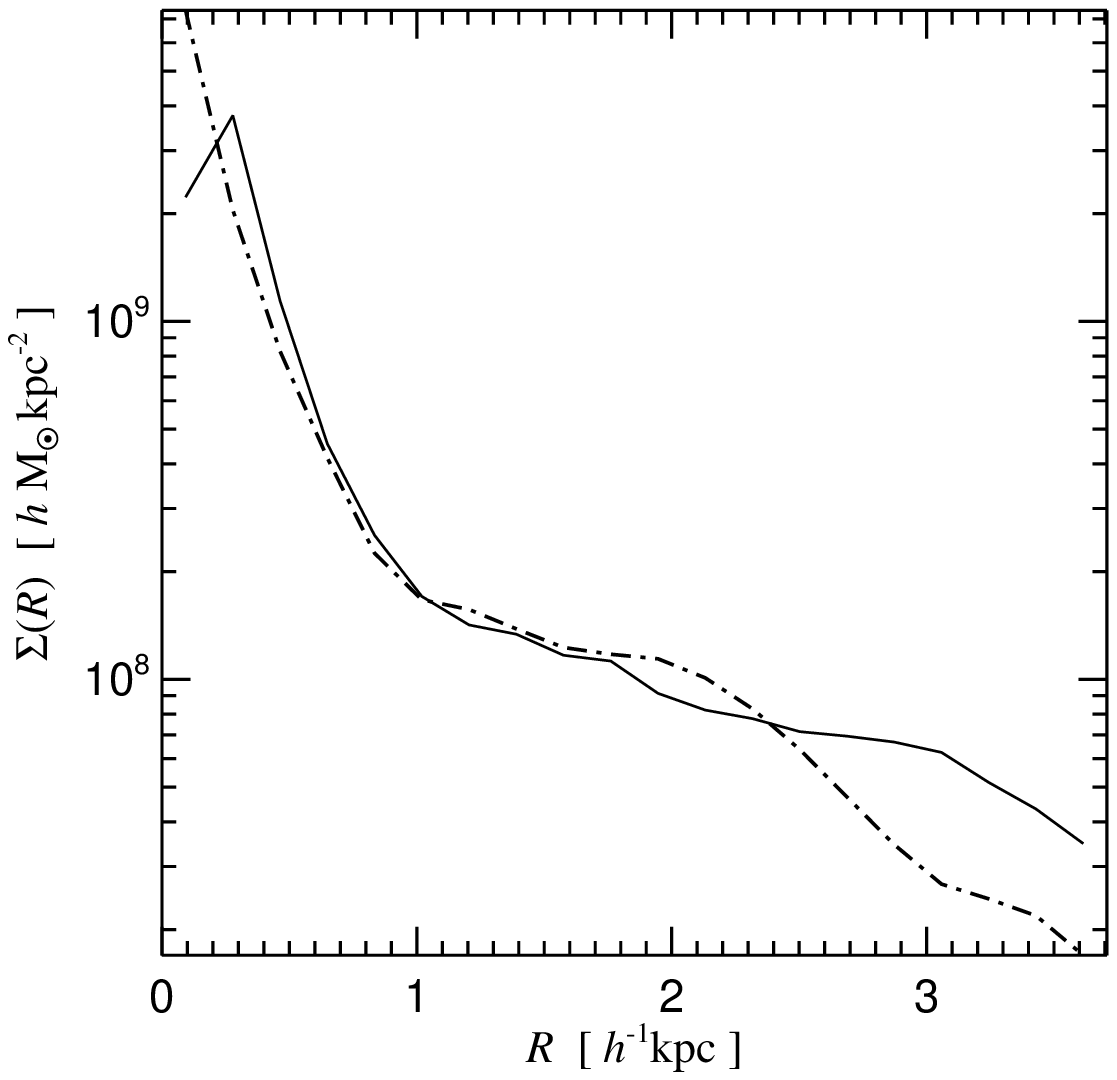,width=\textwidth}   
    \psfig{figure=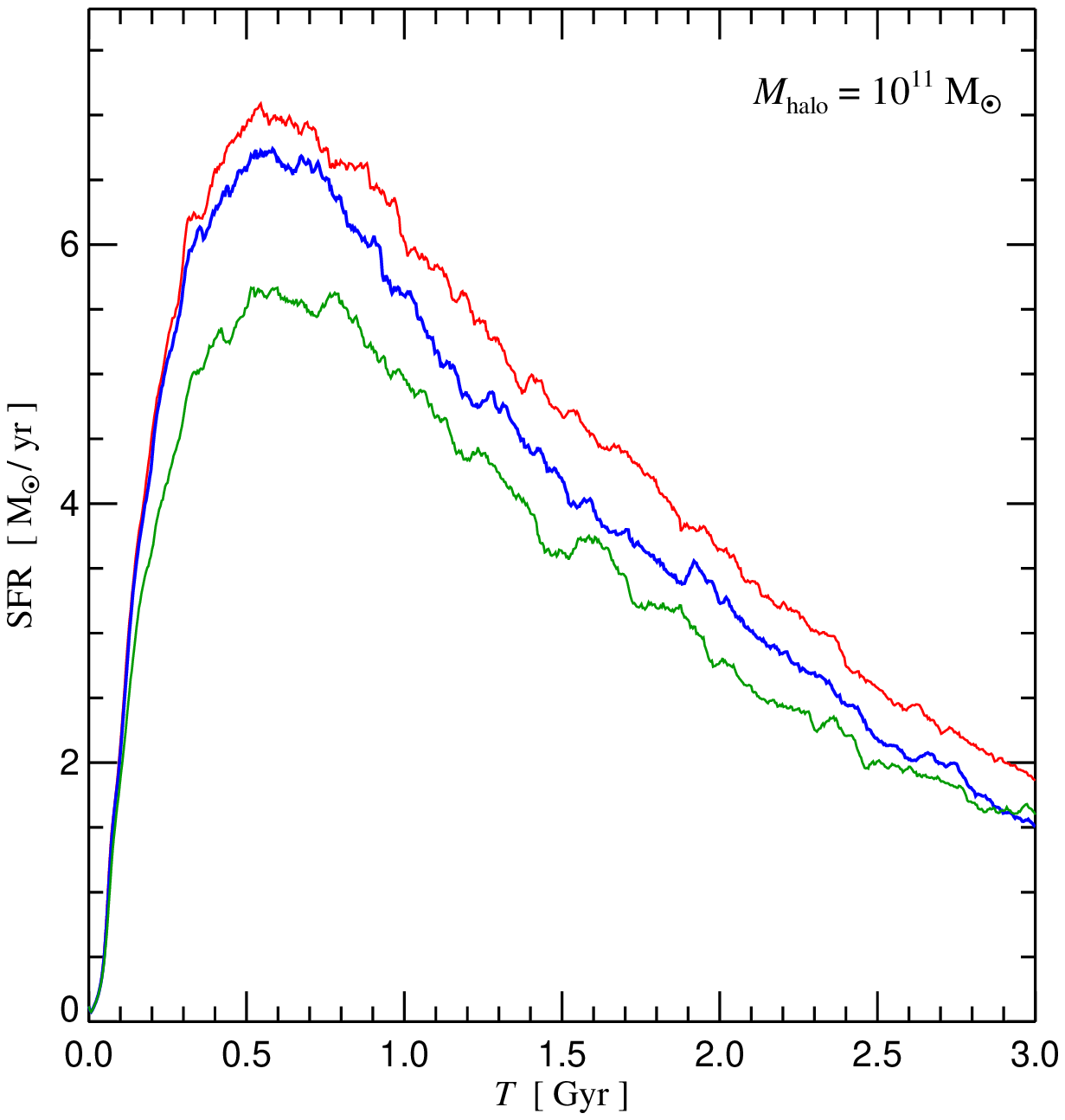,width=\textwidth}     
  \end{minipage}
  \begin{minipage}[t]{0.32\textwidth}
    \begin{center}
    $10^{12} \, M_\odot$      
    \end{center}
    \psfig{figure=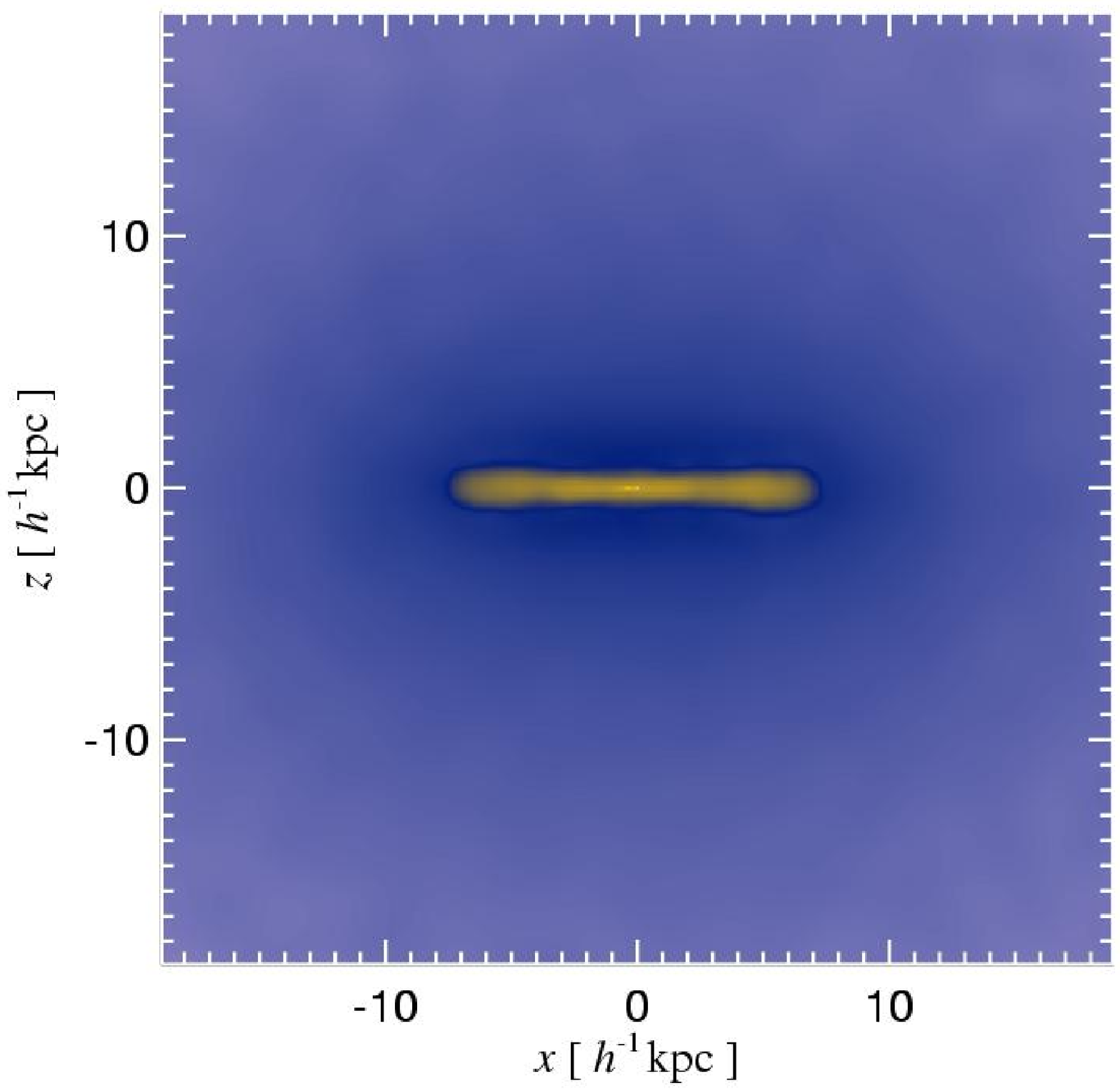,width=\textwidth}    
    \psfig{figure=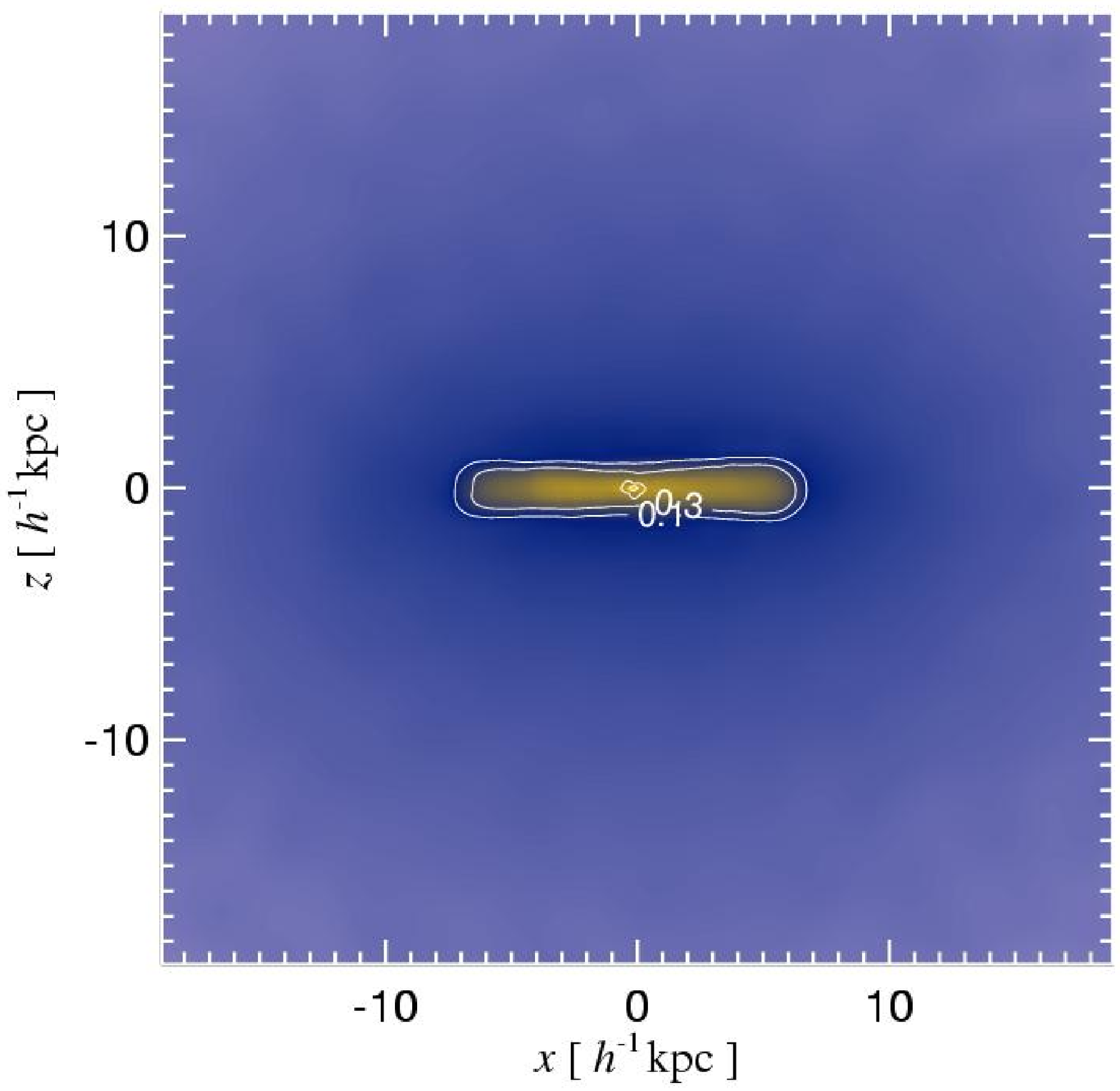,width=\textwidth}    
    \psfig{figure=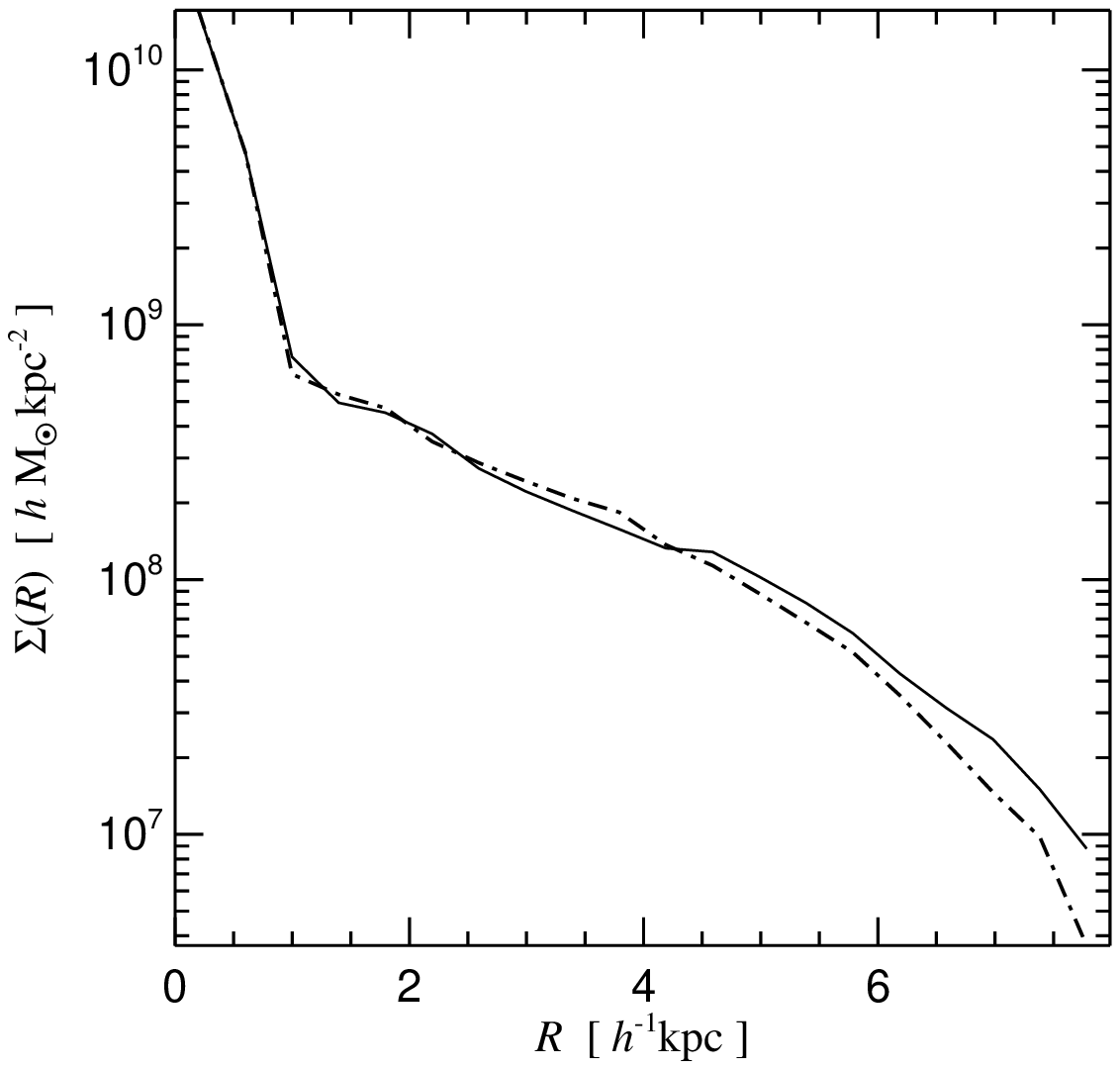,width=\textwidth}    
    \psfig{figure=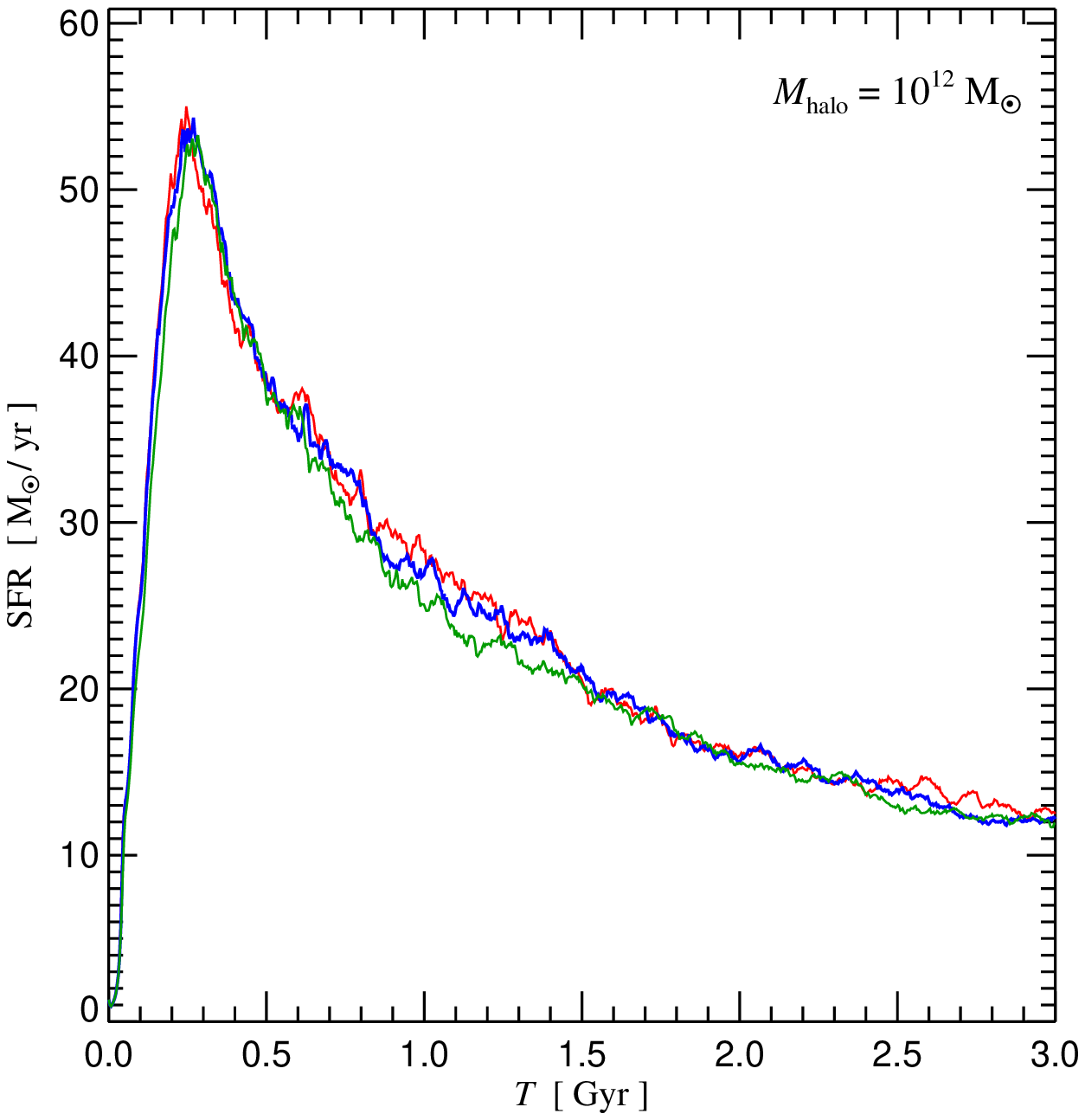,width=\textwidth}     
  \end{minipage}
\caption{Simulation of isolated galaxies with different masses (different
  columns). The top row shows the gas distribution of galaxies without
  CRs. The second row shows the same with CRs included. The contour lines
  indicate the relative level of CR pressure support. Massive galaxies are
  mostly unaffected by CRs, whereas small galaxies exhibit a puffed-up gas
  distribution, and a strongly reduced stellar surface brightness profiles
  (third row). Finally, the star formation histories of galaxies for different
  levels of CR injection efficiencies are shown in the last
  row.}\label{fig:gal}
  \end{figure}

Using a new numerical implementation of self-consistent CR physics in the
parallel SPH code {\sc gadget-2} \citep{Springel, Jubelgas, Pfrommer_a}, we
perform simulations of forming galaxies of different masses, with and without
CR physics. Fig.~\ref{fig:gal} presents the result of our simulation runs. The
morphology and star formation rate of small mass galaxies is strongly affected
by the presence of CRs, whereas massive galaxies appear to be unmodified. The
suppression of star formation by CRs in small galaxies is an attractive
explanation of the observed shallow slope at the faint end of the luminosity
function of galaxies. This suppression, and also the oscillations in the star
formation rate are a result of an inverted effective equation of state of a CR
gas energized by supernovae \citep[see][for details]{Jubelgas}.

\begin{figure}
  \begin{minipage}[t]{0.5\textwidth}
    \centering{Relative CR pressure, \\non-radiative simulation:}
    \psfig{width=\textwidth, figure=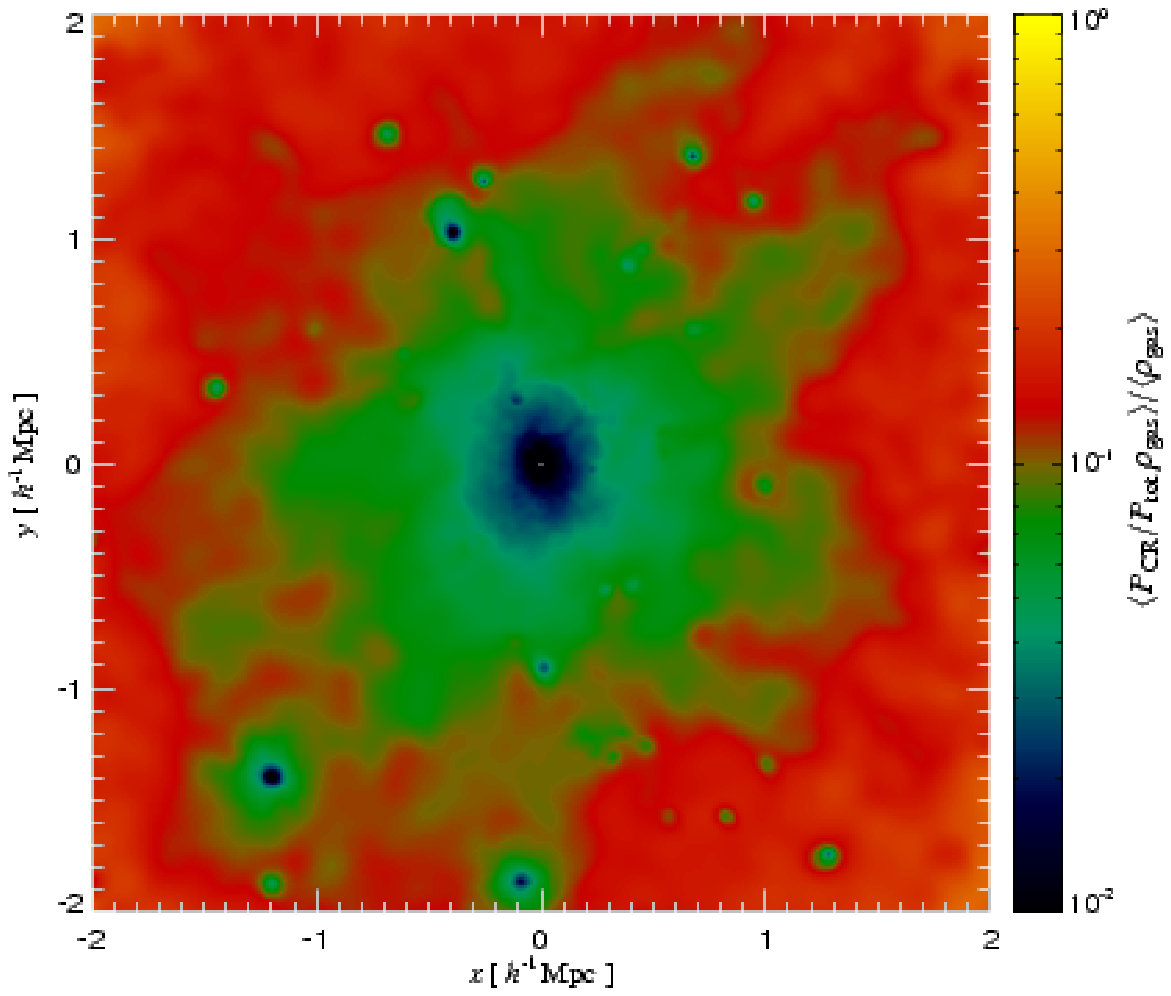}
    \centering{$S_X$ difference map:}
    \psfig{width=\textwidth, figure=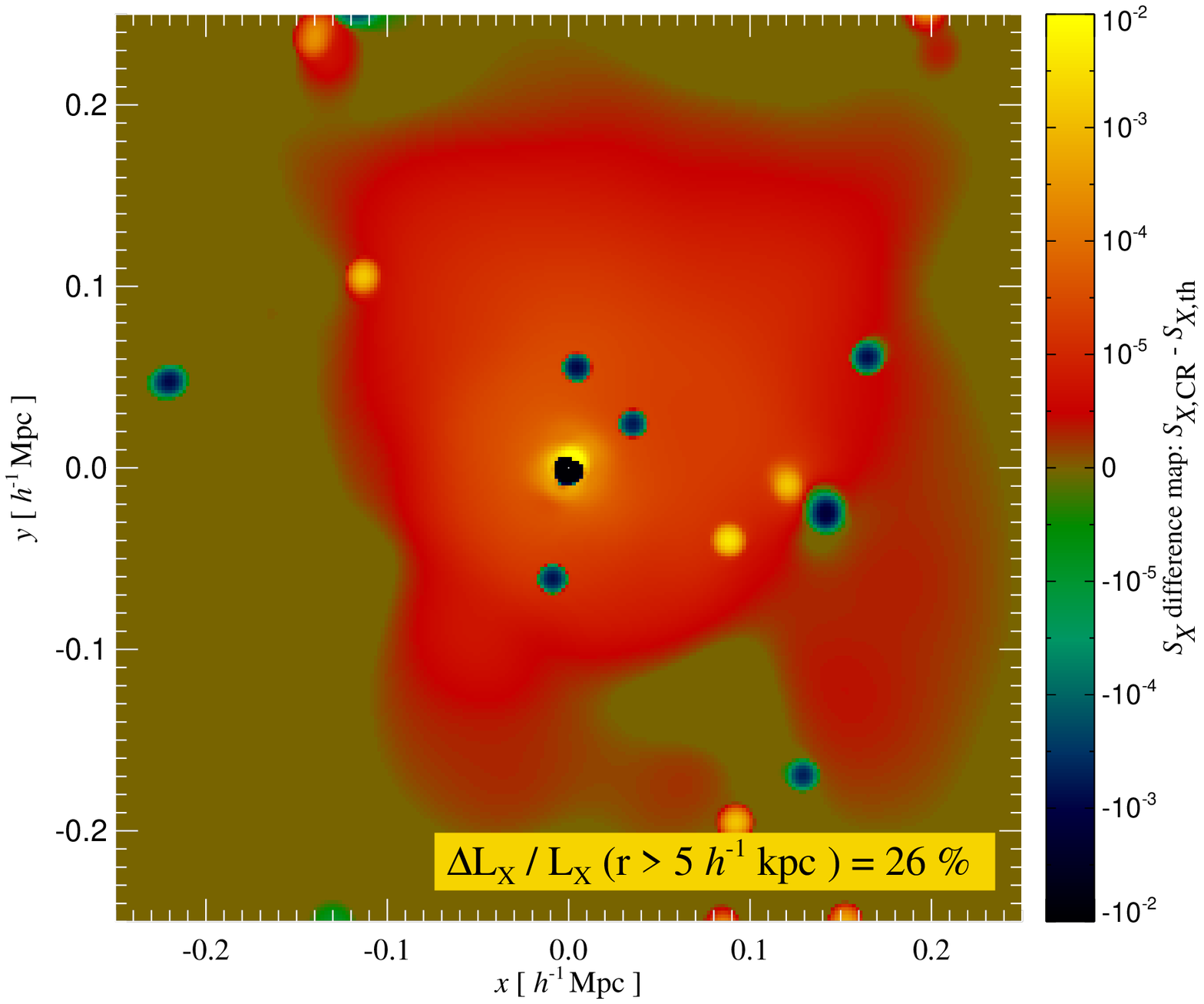}
  \end{minipage}
  \begin{minipage}[t]{0.5\textwidth}
    \centering{Relative CR pressure, \\radiative simulation:}
    \psfig{width=\textwidth, figure=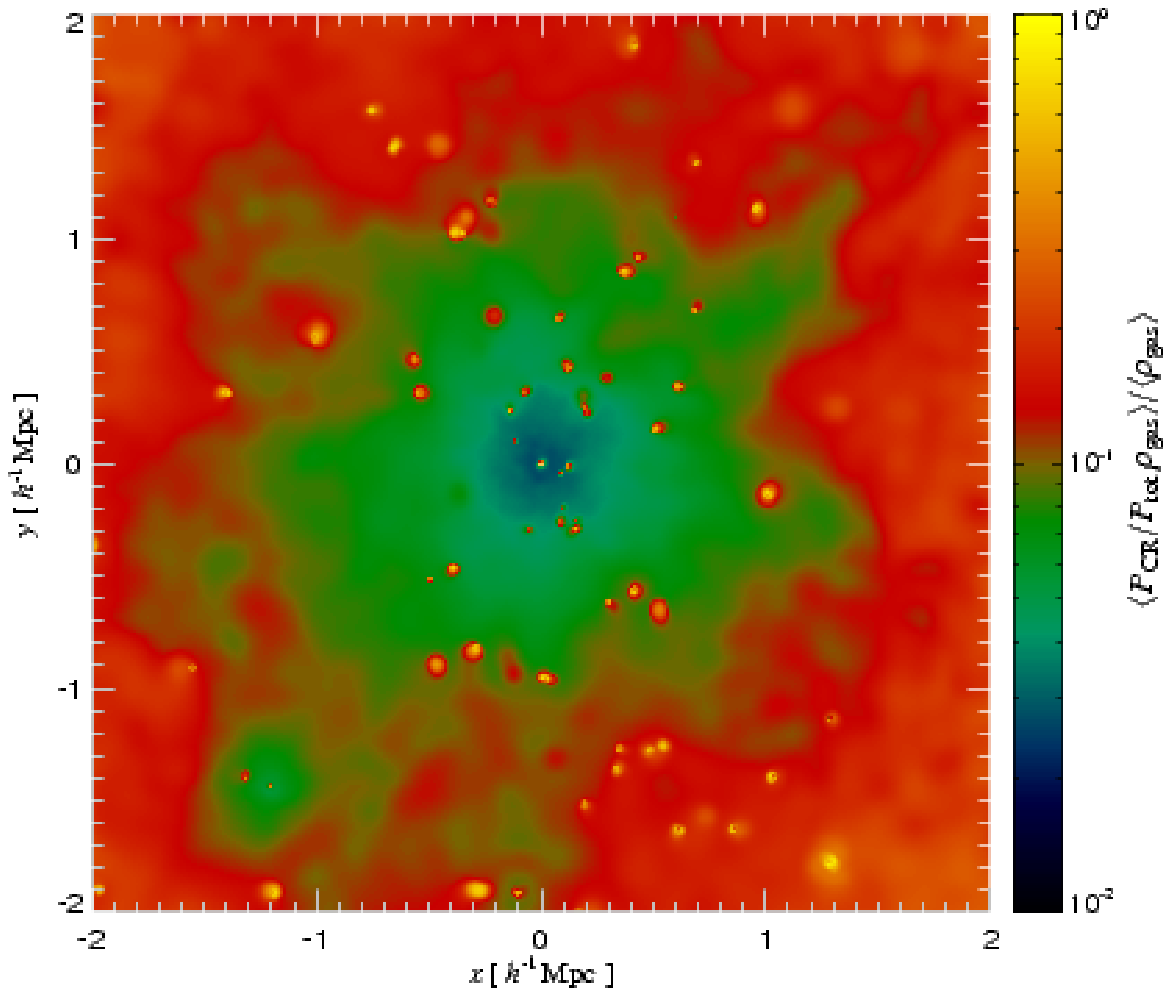}
    \centering{Compton-$y$ difference map:}
    \psfig{width=\textwidth, figure=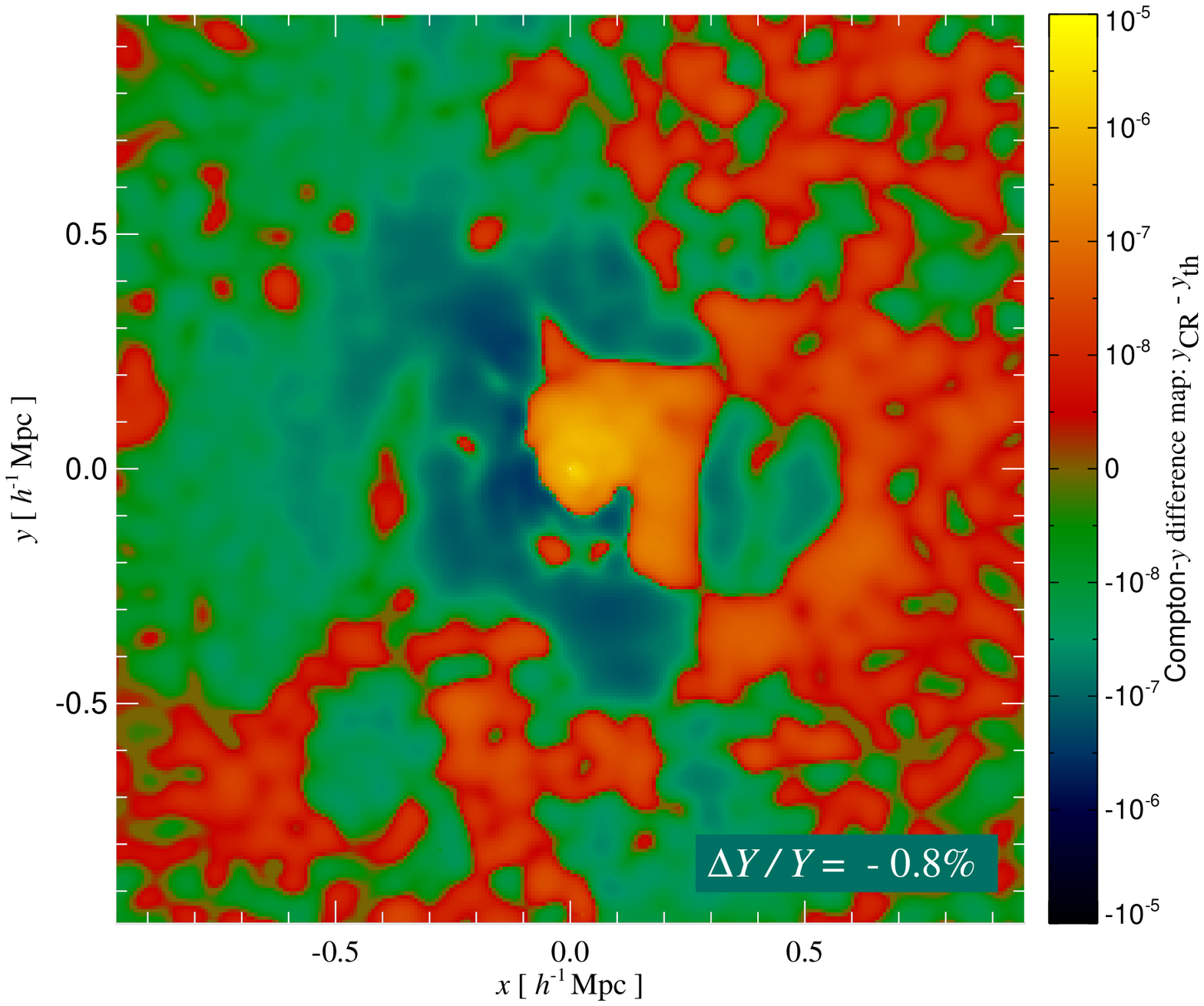}
  \end{minipage}
\caption{The top panels show a visualization of the pressure contained in CRs
  relative to the total pressure $X_\CR = P_\CR / (P_\CR + P_\therm)$ in a
  zoomed simulation of an individual galaxy cluster with mass $M = 10^{14}
  h^{-1} M_\odot$. The map on the {\em left-hand side} shows a non-radiative
  simulation with CRs accelerated at structure formation shock waves while the
  map on the {\em right-hand side} is from a simulation with dissipative gas
  physics including cooling, star formation, supernova feedback, and structure
  formation CRs. The lower panels show the CR-induced difference of the X-ray
  surface brightness $S_X$ ({\em left-hand side}) and the Compton-$y$ parameter
  ({\em right-hand side}) in a radiative simulation with structure formation
  CRs compared to the corresponding reference simulation without CRs. The
  relative difference of the integrated X-ray surface brightness/Compton-$y$
  parameter is given in the inlay. Within cool core regions, the CR pressure
  reaches equipartition with the thermal pressure, an effect that increases the
  compressibility of the central intracluster medium and thus the central
  density and pressure of the gas. This boosts the X-ray luminosity of the
  cluster and the central Sunyaev-Zel'dovich decrement while the integrated
  Sunyaev-Zel'dovich effect remains largely unaffected. }
\label{fig:XCR}
\end{figure}

To study the impact of CRs on larger scales, we performed cosmological
high-resolution simulations of a sample of galaxy clusters spanning a large
range in mass and dynamical states, with and without CR physics. We account for
CR acceleration at structure formation shocks and various CR loss processes
\citep[see][for details]{Pfrommer_b}.  Within clusters, the relative CR
pressure $X_\CR = P_\CR / (P_\CR + P_\therm)$ declines towards a low central
value of $X_\CR\simeq 10^{-4}$ in non-radiative simulations due to a
combination of the following effects: CR acceleration is more efficient at the
peripheral strong accretion shocks compared to weak central flow shocks and
adiabatic compression of a composite of CRs and thermal gas disfavours the CR
pressure relative to the thermal pressure due to the softer equation of state
of CRs. Interestingly, $X_\CR$ reaches high values at the centre of the parent
halo and each galactic substructure in our radiative simulation due to the fast
thermal cooling of gas which diminishes the thermal pressure support relative
to that in CRs.  This additional CR pressure support has important consequences
for the thermal gas distribution at cluster centres and alters the X-ray
emission and the Sunyaev-Zel'dovich effect (cf. Fig.~\ref{fig:XCR}).

\vspace{-0.25cm}
\section{Conclusions}

We have argued that cosmic rays play an active role for galaxies and the large
scale structure. We find that star formation is strongly CR-suppressed in small
galaxies. This is a result of the modified effective equation of state of CRs
in the ISM and leads to a suppression of the faint end of the galaxy luminosity
function and thereby helps to reconcile observations with computational models
of galaxy formation. In galaxy clusters, the X-ray luminosity is boosted
predominantly in low-mass cool core clusters due to the large CR pressure
contribution in the centre that leads to a higher compressibility. The
integrated Sunyaev-Zel'dovich effect is only slightly changed while the central
flux decrement is also increased.


\vspace{-0.15cm}
\begin{thebibliography}{}
\bibitem[En{\ss}lin et al.(2006)]{Ensslin}En{\ss}lin, T.A., Pfrommer, C.,
  Springel, V., \& Jubelgas, M. 2006, arXiv:astro-ph/0603484

\bibitem[Jubelgas et al.(2006)]{Jubelgas}Jubelgas, M., Springel, V., En{\ss}lin,
  T.A., \& Pfrommer, C. 2006, arXiv:astro-ph/0603485

\bibitem[Pfrommer et al.(2006a)]{Pfrommer_a}Pfrommer, C., Springel, V., En{\ss}lin, \&
  Jubelgas, M. 2006, MNRAS, 367, 113

\bibitem[Pfrommer et al.(2006b)]{Pfrommer_b}Pfrommer, C., En{\ss}lin, Springel, V., 
  Jubelgas, M., \& Dolag, K. 2006, submitted, arXiv:astro-ph/0611037

\bibitem[Springel(2005)]{Springel}Springel, V. 2006, MNRAS, 364, 1105
\end{thebibliography}
\end{document}